\documentclass{llncs}
\usepackage{llncsdoc}
\usepackage{graphicx}
\usepackage{amssymb}
\begin{document}
\title{A formal definition and a new security mechanism of physical unclonable functions}

\author{Rainer Plaga \and Frank Koob}

\institute{Federal Office for Information Security (BSI), D-53175 Bonn, Germany\\
$\{$rainer.plaga,frank.koob$\}$@bsi.bund.de}

\maketitle
\begin{abstract}
The characteristic novelty of what is generally
meant by a ``physical unclonable function'' (PUF)
is precisely defined, in order
to supply a firm basis for security evaluations
and the proposal of new security mechanisms.
A PUF is 
defined as a hardware device which
implements a physical function with an
output value that changes with its argument.
A PUF can be clonable, but a secure PUF must be
unclonable.
\\
This proposed meaning
of a PUF is cleanly delineated from
the closely related concepts
of ``conventional unclonable function'',
``physically obfuscated key'',``random-number generator'',
``controlled PUF'' and ``strong PUF''.
The structure of a systematic security evaluation of a PUF enabled by the proposed
formal definition is outlined.
Practically all current and novel physical 
(but not conventional) unclonable physical functions 
are PUFs by our definition.
Thereby the proposed definition
captures the existing intuition about what is a PUF and remains
flexible enough to encompass further research.
\\
In a second part we quantitatively characterize two classes of PUF security mechanisms,
the standard one, based on a minimum secret read-out time,
and a novel one, based on challenge-dependent erasure of stored
information. 
The new mechanism is shown to allow in principle the construction
of a ``quantum-PUF'', that is absolutely secure while not requiring the
storage of an exponentially large secret. The construction
of a PUF that is mathematically and physically unclonable in principle
does not contradict the laws of physics.
\end{abstract}
\section{Introduction}
\subsection{Aims and outline of this work}
``Physical unclonable functions'' (PUFs) are electronic hardware devices that 
are hard to reproduce and can be uniquely identified
\cite{pappu,gassend04}.
They promise to enable qualitatively novel security mechanisms
(see e.g. \cite{armknecht2,gassend08,sram})
and have consequently become a ``hot topic'' in hardware security\cite{busch}.
The present work asks  
the question ``What characteristics exactly define the qualitative novelty
of the PUF concept?''. We hope that a precise answer will aid the
security evaluation of existing PUFs and help to develop new ideas for PUF security mechanisms.
We searched for
\begin{enumerate}
\item a formal definition of the properties that are required from a hardware device to be called ``PUF'',
and a
\item a formal definition of the criteria that have to be fulfilled to consider a PUF ``unclonable''.
\end{enumerate}
The formal PUF definition should
not suffer from weaknesses of previous definitions (see section
\ref{previ}),
encompass at least the large majority
of the existing PUF constructions, and be as flexible as possible,
i.e. does not restrict further
progress in PUF development (e.g. by demanding constructional
details, like the amount of stored information). This
aim is achieved in section \ref{form}.
After formulating a simple definition of PUF-security (based
on Armknecht et al.\cite{armknecht}) in section \ref{pufsec}
we delineate PUFs from some closely related security concepts
(section \ref{related}) and outline the elements
of a PUF-security evaluation (section \ref{security}).
In a second part of the paper 
we systematically analyse and classify PUF security mechanisms and calculate
their quantitative security levels against attacks that attempt mathematical
cloning (section \ref{prac}).
The aims of this section are to give a quantitative answer 
to Maes \& Verbauwhede's\cite{maes} question whether
mathematically-unclonable PUFs are possible in principle,
and to apply and thereby illustrate the PUF-definitions
of the first part of the paper.
In section \ref{discu} we 
characterise the qualitative novelty of PUFs 
as a new primitive of physical cryptography
and discuss the future use 
and development of PUFs.

\subsection{Previous work on the definition of a PUF}
\label{previ}
There have already been several proposal for the first definition of required PUF properties.  
Gassend et al.\cite{gassend04} 
who invented the term ``PUF'' (earlier work by Pappu was
on the slightly different concept of a physical one-way function\cite{pappu})
demand that the function must be ``easy to evaluate'', i.e. it must efficiently
yield a response value ``R'' for a challenge argument ``C''.
and ``hard to predict (characterize)''. The latter property means
that an attacker who has obtained a polynomial number of C -- R pairs (CRPs) but 
has no longer physical access to the PUF can only extract a negligible 
amount of information about the R for a random C.
R\"uhrmair et al.\cite{ruehr2} criticised this definition because the
information content of finite physical objects is always
polynomially bound, and therefore no PUF fulfilling this
definition can exist. They propose an alternative formal
definition in which the PUF must only be hard to predict 
for an attacker ``who may execute any physical operation
allowed by the current stage of technology''. 
Maes \& Verbauwhede\cite{maes} chose to {\it exclude} unpredictability
from their ``least common property subset'' of PUFs, because they
put into question whether it is possible in principle to construct a
mathematically unclonable PUF. They demand
that a PUF is ``easy to evaluate'' (property
``evaluatable'') and that it is ``reproducible'', meaning that a
C always leads to the same R within a small error.
Moreover they demand ``physical unclonability'' i.e. that
it must be ``hard'' for an attacker to construct a device that
reproduces the behaviour of the PUF.
However, PUFs that are mathematically clonable are
also physically clonable because the mathematical
algorithm for PF can then be implemented on a device
that is then a functional physical clone of the PUF.
Summarizing, a first generation of definitions
roughly defined PUFs to be devices that are efficiently 
evaluatable 
and are mathematically and physically unclonable. 
They remain unsatisfactory for two reasons:
\begin{enumerate}
 \item Most of the devices currently called PUFs do not
fulfill these definitions (according to R\"uhrmair et al.\cite{ruehr2} there
are only some ``candidates''), i.e. the definition does evidently not
really capture the PUF concept.

 \item They combine the definition of a PUF with the
definition of its security, i.e. points 1. and 2. above.
A PUF is defined by its unclonability i.e. its security against attacks.
This is problematic because an open-ended security analysis of a PUF
clearly must have an ``insecure PUF'' as one a priori possible outcome.
Based on the above definitions an ``insecure PUF'' is a paradox, PUFs
would be secure by definition.
\end{enumerate}
These two problems were elegantly solved 
in a seminal paper by
Armknecht et al.\cite{armknecht} who propose to
formalize a PUF as ``physical function (``PF'') - which is
a physical device that
maps bit-string-challenges ``C''
to bit-string-responses ``R''.
The unclonability is recognized by 
Armknecht et al. as only one crucial security property, that
they further formally define in great detail.
We will supply a simplified version of their
general security definition in section \ref{pufsec} below.
Following Armknecht et al., the PUF definition 1. consists in an
answer to the question: What are the required 
characteristics of PF() in
order to be a PUF? Armknecht et al. do not
demand any specific mathematical properties but only
that a PF is a ``probabilistic procedure'' that maps
a set of challenges to a set of responses and
that internally PF is a combination of a
physical component and an evaluation
procedure that creates a response. 
Armknecht et al. explain that the responses rely heavily on the
properties of the physical component but also 
on uncontrollable random noise 
(hence ``probabilistic'').
This definition of PF() still faces the following
problem:
\begin{itemize}

\item
Consider a standard authentication chip
with a stored secret in a physically protected memory
that calculates a response from the challenge
and the secret. Such a chip must contain
a ``physical component'' (the memory)
and an evaluation procedure (its read-out)
that fulfills Armknecht et al.'s definition because
some (very small) uncontrollable random noise
is unavoidable even in standard computer
memories. There is also no reason
why a well designed standard authentication chip
cannot posess Armknecht et al.'s
security properties.
\end{itemize}
Therefore,
even though Armknecht et al.'s definitions
constitute great progress of lasting value,
they still do not capture 
the distinctive properties of the PUF concept.
In practice Armknecht et al. define all devices 
that run any challenge-response protocol as PUFs.

\section{A model of the PUF concept}
\subsection{Formal definition of ``PUF''}
\label{form}
\begin{figure}
\includegraphics[angle=0,totalheight=2.5in]{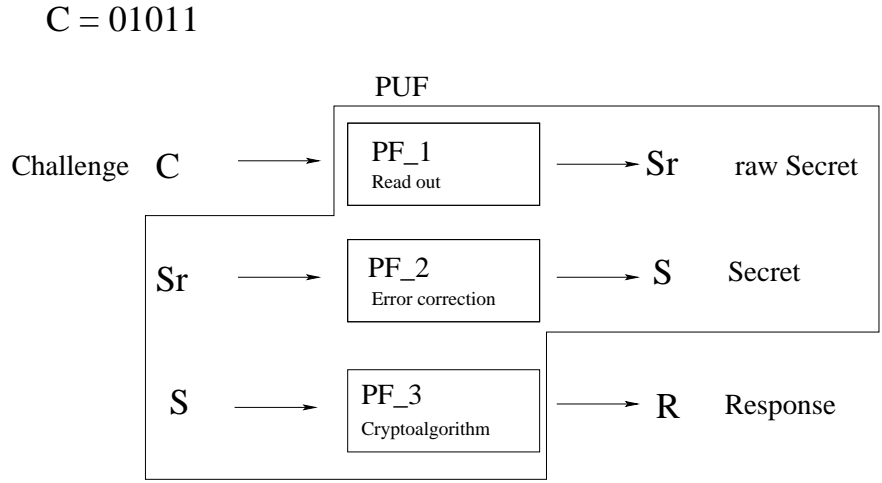}
\caption{
{\bf Symbolic model of a PUF} The box delineates the PUF
that receives a challenge C (shown with an
example bit string) and sends a response R that is determined in
three distinct steps. The first step is the physical readout,
the second the correction of errors that can occur in the first
step and the third step includes all operations of mathematical
cryptography.}
\label{explot}
\end{figure}

In the following we assume Armknecht et al.'s 
model of a PUF as physical function PF() (see section \ref{previ}).
We break up the physical function PF() into
three steps  (see fig.(\ref{explot})). C,S$_r$,S and R are bit strings.
\begin{enumerate}
 \item 
In the first ``physical read-out'' step PF$_1$ = S$_r$, internal information S$_r$
(the ``raw secret'') is physically read-out from the PUF in response 
to a challenge C foreseen by the system architecture.
\item
In an optional second step PF$_2$(S$_r$) = S error correction and/or 
privacy amplification are performed, such that errors in the
read-out are corrected and parts of S$_r$ which may be
known by the attacker (e.g. by guessing parts of the
challenge) are removed by privacy amplification algorithms.
\item
In an optional third step PF$_3$(S) = R,
some additional algorithm is performed with
S as input
to calculate the final response R. Typically PF$_3$ is 
some cryptographic protocol that proves the possession of
S without disclosing it. 
\end{enumerate}
In many existing PUF architectures the challenge C is 
an address of information
inside the PUF which is output as the response R. 
E.g. in arbiter PUFs\cite{lim} C defines the choice of a set of delay switches
whose cumulative delay path defines S (and from this R).
Our idea is that the possibility for this
mode of addressing, rather
than its ``unclonability'', defines a PUF. 
The challenge C can then be understood as a 
key required for physical access to the response R. R remains secret
without access to C.
\\
A security architecture based on this idea requires
certain properties of PF() which define the PUF
concept:
\begin{itemize}
 \item 
 {\bf Formal Definition 1 of a PUF} 
{\it A hardware device is called ``PUF'' if: 
\\
a. a physical function PF$_2$(PF$_1$()) 
which is deterministic for a specific
set of challenges $\mathfrak{M}$, can be evaluated with each challenge at least
once and
\\
b. the value S = PF$_2$(PF$_1$(C))
changes with its argument, for all outside challenges C $\in$ $\mathfrak{M}$,
i.e. PF$_2$(PF$_1$(C)) = S is not
a constant function.}
\end{itemize}
One difference to some previous PUF definitions is that PF()
is not required to be {\it easily} evaluatable. An 
efficient evaluation of S is certainly
a desirable design goal, but there is no reason why 
a device with inefficient read out cannot be a PUF
by definition.
\\
Another difference to most previous definitions is 
that it allows a PUF to be clonable.
As an example consider the following physical function that fulfills
the above definition 1:
\begin{itemize}
\item
 PF$_1$(any C with more 1s than 0s) = 1001101101
\item
 PF$_1$(any C with more 0s than 1s or equal number of
1s and 0s) = 0001101000
\end{itemize}
Clearly a PUF with this PF$_1$ can be reproduced by
a trivial algorithm, i.e. it is trivially mathematically clonable.
This is a desirable property because 
``clonable PUFs'' do exist in the real world and
should not present a PUF definition with a paradox.
``Unclonability'' is then a property that is aimed for,
rather than achieved by definition. Analogously ``cryptography'' aims for
secrecy (crypto) rather than achieving it by definition.
Even though it is a child's game to break
it, the Cesar cipher is a valid cryptographical
algorithm according to this definition.
Consequently
cryptographic algorithms are commonly defined to be ``key-dependent
injective'' (rather than ``unbreakable'') mappings''\cite{crypto}.
\\
Where does this leave PUF security?
It is not possible in principle to extract the secret S
from a PUF without knowing of the challenge.
This is true even for the above insecure PUF. 
However in the example above it is easy to reproduce PF$_1$, and therefore,
as soon as the challenge becomes known, S becomes known. Therefore
the crucial necessary objective for the 
security of a PUF is the unclonability of PF$_2$(PF$_1$()).
In the next section we make this insight more precise.
A complete and quantitative set of security requirements
(i.e. with requirements on their length $\ell$, the number of independent
challenges N etc.) can only be made in the context of a concrete PUF architecture.
One example is discussed further in section \ref{mrt}.

\subsection{Security of a PUF}
\label{pufsec}
\subsubsection{Requirements for prediction}

Even though the response of a PUF can in principle be used for
various purposes, we will conclude
in section \ref{discu} that one central 
PUF capability is the distribution of remote authentication secrets.
If S is used for authentication
purposes, an attacker must be able to fully predict it, i.e. a partial
prediction of S=PF$_2$(PF$_1$(C)) for a given argument C 
will not be considered a successful attack
in the following. Therefore, the natural ``basic objective'' 
of PUF security is that the attacker cannot predict a complete, correct
bit string S for a given bit string C.
\subsubsection{Attack models}
\label{attack}
Security can only be defined relative to an attack model, that
lays down the assumptions about the security environment.
We assume in the following
two models from the literature that seem realistic
in practice. The first one models an attempt to break
Armknecht et al.'s\cite{armknecht} selective 
unclonability\footnote{R\"uhrmair et al.'s\cite{ruehr2} 
PUF definition demanded
that the original manufacturer of the PUF cannot 
produce two PUFs which are clones of each other (Armknecht et al.\cite{armknecht} demand this
``existential unclonabiliy'' 
only optionally.). 
``Selectively unclonability''\cite{armknecht} means that given physical access to the device an attacker
cannot produce a clone. 
\\
In practice existential unclonability would hardly
enhance the security against a malicious manufacturer, for the following reason. 
She could produce ``quasi-existential-PUF'' devices that do not
meet the PUF definition 1, but algorithmically
simulate - e.g. with an keyed hash function - an output that cannot be discriminated from 
the one of an existential PUF. These quasi-existential-PUFs could be easily cloned
by the malicious manufacturer, and could serve exactly the same purpose as
clonable PUFs.
\\
As an alternative  to existential
unclonability we will propose a weaker
``resistance-against-insider-attacks''
security level in this section \ref{attack}.}.
It does not put any restriction on the attack
strategy, therefore adaptive choices of challenges
are possible\footnote{Therefore strong unpredictability 
in the sense of Armknecht et al.\cite{armknecht} will be necessary 
to protect the PUF.}.
The second one is an attempt to do the same with
a certain reasonable amount of insider knowledge.
Both models assume that the attacker has only
access to one single PUF, i.e. attacks exploiting correlations
between different PFs are excluded by assumption (see Armknecht
et al.\cite{armknecht} for the general case).
\\
{\it Attack model 1: ``Outsider attack'': 
The attacker has physical access only to the attacked
PUF only for a finite amount of time $\Delta$t$_a$. 
After this access period,
she tries to predict a secret S from the PUF to a challenge C, randomly 
chosen from the set of all challenges.
She has no knowledge
of the set of challenges and secrets that will be used during the
active lifetime of the PUF or any further
previous knowledge of the PUF.}
\\
{\it Attack model 2: ``Insider attack'': 
The attacker has physical access only to the attacked
PUF only for a finite amount of time $\Delta$t$_a$.
After this access period 
she tries to predict a secret S from the PUF to a challenge C, randomly 
chosen from the set of all challenges.
She has no knowledge
of the set of challenges and secrets that will be used during the
active lifetime of the PUF 
but she has all other information 
that the manufacturer of the PUF has about the attacked
individual PUF.}
\\
The attack models assume that the attacker tries to predict S 
rather than R, because PF$_3$ might be protected with non-PUF security mechanisms,
e.g. with a secure tamper-resistance scheme in combination with a
secure crypto algorithm. Such a security
mechanism shall remain out of our consideration because we 
aim to define the security of
the PUF itself.
\\
Security against a model-2 attacker
corresponds to unclonability against an
attacker who has most of the inside knowledge about 
the PUF production, but who cannot directly manipulate
the production process. This unclonability is weaker
than ``existential unclonability'' (see footnote 1) but perhaps more relevant
in practice.
\subsubsection{Definition of a secure PUF}
\label{securepuf}
The PUF-security definition now follows from the
requirement that the attack shall be unsuccessful:
\begin{itemize}
\item
{\bf Formal definition 2 of the PUF-security objective}
\\
{\it A PUF is secure against an attack of a model-1 
(``selectively unclonable'') attacker if a
model-1 attacker
can compute or physically copy the function
PF$_2$(PF$_1$(C)) = S
for not more than a negligible fraction L of challenges from 
the set of all possible challenges.
Here ``compute'' means via a computation independent
of the PUF and corresponds to ``mathematical cloning''. 
``Physically copy'' means to create a device that functionally reproduces 
PF$_2$(PF$_1$(C)) and corresponds to ``physical cloning''.}
\end{itemize}
Replacing the model-1 by a model-2 attacker defines
a PUF that is ``insider selectively unclonable''.
L is the security level of a secure PUF, i.e.
the probability for an attacker to successfully predict the secret
S for a challenge C without being in posession of the PUF after
the access period.
\\
A precise quantification of ``negligible'', i.e. the decision which upper limit
of L is required, cannot be made on the level of this general definition
because it depends on the detailed
security environment. 
L is analogous to the required probability p of
a successful brute force attack in classical cryptography
that depends on the key length.
We propose as a reasonable upper limit on L that it is ``negligible on a terrestrial
scale'' which has been estimated by
Emile Borel as $<$ 10$^{-15}$\cite{borel}.

\section{Relation of PUFs to closely related concepts}
\label{related}
In this section we delineate the concept of a PUF as defined
in section \ref{form} and \ref{securepuf} from
five closely related concepts.

\subsection{PUFs and conventional unclonable functions (``CUFs'') are qualitatively different}
\label{cuf}
Let us first differentiate between a  PUF and a conventional
physical function that serves the same function as a PUF
(called ``conventional unclonable function'' CUF in the following).
A CUF contains secret information that is protected 
by tamper resistance, by anti side-channel- and
fault-induction-attack measures and by a cryptographic
algorithm that protects the secret from
disclosure via the response.
A CUF does not fulfill the PUF definition 1., because the secret
does not depend on the challenge. In other words:
The first physical secret readout step PF$_1$(C) is a constant
function in a CUF. 
\\
PUF and CUF differ qualitatively in the way
they protect the secret. In a PUF
the lack of knowledge of the challenges protects the secret S 
in a similar sense that the
lack of knowledge of a cryptographical
key protects the clear text in a cipher text. There is no
analogous ``key'' in a CUF.
Its security mechanisms merely rely
on physical barriers and arrangements that
prevent access to secret information.

\subsection{PUFs and physically obfuscated keys are independent concepts}
\label{pok}
Devices that
extract physical information 
with ``non-standard'' 
methods are currently called PUF even if there is no (or effectively a single fixed internal)
challenge (e.g. in SRAM PUFs\cite{sram}).
In this case PF$_1$() is formally constant, so that 
such devices are no PUFs in the sense of our definition 1.
We endorse R\"uhrmair et al.'s suggestion\cite{ruehr2} to call 
information extracted in this way in general ``physically obfuscated keys'' (POKs).
This limit of N=1 is the only one where devices that are currently 
called PUFs, would no longer be classified as PUF under our proposed definition.
We find this appropriate because while 
POKs can enable valuable tamper-resistance mechanisms (see below), they 
are not the {\it qualitatively} novel primitive of physical cryptography
that PUFs promise to be 
(see section \ref{discu} for further discussion of the nature of this primitive).
\\
The protection by obfuscation is valuable: 
it consists in the extra-time an attacker needs to
learn the non-standard readout mechanism or position
in a standard memory 
where an obfuscated key has to be stored at least temporarily.
POKs are delineated from CUF only by the ``non-standard'' qualifier because  
stored information is {\it always} physical\cite{landauer}.
The secrets of PUFs will usually be stored in a non-standard way, i.e.
they will also be POKs.
But this is no necessary requirement for a PUF. There is no fundamental reason why
PUFs cannot have ``standard'' computer memories (see e.g. SHICs\cite{shic}, a PUF
using a standard crossbar memory).
\\
Physically obfuscated functions (POFs) may also appear in 
PUF architectures. They are defined as computation with non-standard
physical processes, e.g. via scattering of light or folding of proteins.

\subsection{Random number generators}

In both deterministic and physical random number generators
the initial read-out step PF$_1$ (the read out of the seed) does not depend
on a challenge C. In secure deterministic RNGs PF$_1$(C)
must be a constant function. In physical RNGs PF$_1$ is not
constant but intrinsically random, i.e. not deterministic. 
Therefore, RNGs do not meet the PUF definition 1.

\subsection{Controlled PUFs: a PUF with additional tamper
resistance}

In controlled PUFs\cite{cpuf,gassend08} tamper-resistance measures prevent
the attacker from obtaining C -- S$_r$ pairs from the PUF.
Only the C -- R pairs - from which S$_r$ cannot be derived
if PF$_3$ is a suitable, secure cryptographical algorithm -
can be accessed by an attacker.
It seems likely that PUFs e.g. used in smart cards will eventually all be 
controlled PUFs, because such an additional
well understood security layer stands to reason. However the security of 
PUFs themselves should be analysed under the assumption of
no such a control because if one trusts the control mechanism, 
mathematically clonable PUFs suffice anyway.

\subsection{``Strong PUFs'': not the only path to strength}

R\"uhrmair et al.\cite{ruehr2} defined a PUF to be ``strong'' if
it ``has so many C -- R pairs ... that an attack ... based on
exhaustively measuring the C -- R pairs has a negligible probability
of success''. In our nomenclature a strong PUF is roughly a MRT-PUF 
that fulfills our second security requirement 
(see section \ref{mrt} below, for further explanation of MRT). It is thus appropriate
to call them ''strong``, but there can be secure PUFs which are
not ``strong'' in R\"uhrmair et al.'s sense, e.g. EUR-PUFs(see section \ref{quant}
below for further explanation of EUR).
 


\section{Security evaluation of PUFs}
\label{security}
A main purpose of the present
proposed formal PUF definitions 1. and 2. of the concept ``secure PUF''
is to establish a consistent basis for security evaluations
and certifications of PUFs.
What is the structure of an evaluation on this
basis?
\\
If the proposed PUF
fulfills definition 1,
the basic informal questions of a security evaluation based on 
definition 2 are:
\begin{enumerate}
\item
Which form has PF$_1$(C) and
by which physical mechanism is S$_r$ extracted?
\item
What is the form of PF$_2$(S$_r$)=S and how is the function
evaluated physically?
\item
What is the total information content in the set of all secrets S? 
\item
For what fraction L of the allowed challenges can PF$_2$(PF$_1$(C)) be either
mathematically computed or physically copied?
\item
Which comprehensible physical security mechanisms prevent
an attacker to compute or copy PF$_2$(PF$_1$(C))
for more than a fraction L of all challenges?
\end{enumerate}
Answers to questions 1. - 4. allow to evaluate 
quantitative and comprehensible
security levels against ``mathematical-cloning brute force attacks''
(see section \ref{prac}).
Question 5 will have a more qualitative answer, similar to answers to the question
whether a mathematical cryptographic algorithm is secure
against non-brute force attacks.

\section{Analysis of PUF security mechanisms}
\label{prac}
The holy grail of PUF construction is to construct PUFs
that are unclonable i.e. fulfill the security definition 2 (section \ref{securepuf}).
If an attacker succeeds to access the PUF's internal secrets, she
will usually be able to compute PF$_2$.
Because physical reproduction of a PUF without knowledge
of its internal secrets will probably be hard in practice\footnote{
But not necessarily impossible. She could e.g. succeed to reproduce to clone a PUF
exactly copying its production process.},
PUF security mechanisms must above all
prevent the attacker from {\it computing} PF$_2$.
In other words: mathematical unclonability is the hardest nut.
Therefore we will classify the known PUF security mechanism and
calculate their security level against brute-force mathematical
cloning attacks.
\\
Up to now all proposed and constructed PUFs\footnote{In the sense of this
paper, i.e. excluding POKs.} are based on a mechanism
that we propose to call ``minimum readout time''(MRT) and
that is further discussed
in subsection \ref{mrt}. All these existing PUFs turn out to
fulfill our PUF-definition 1, i.e. they  ``remain'' PUFs, 
even in case they have turned out to be clonable (see below).
Because currently the MRT mechanism dominates the field, one might
be tempted to equate the very
concept of PUFs with it.
However, the flexibility of our definition
allows a completely different PUF security
mechanism that we call ``erasure upon read-out''(EUR) (see section \ref{quant})
for devices.
One concrete EUR PUF, the quantum PUF will be introduced below.
\\
These examples show that our proposed definitions have achieved their
aims: nearly all existing (MRT) PUFs can be included in its
scope, but its flexibility allows to include completely
novel PUF constructions (the EUR PUFs).


\subsection{``Minimum readout time'' PUFs} 
\label{mrt}
This well known PUF security mechanism
is to store a large enough number N of C -- S pairs 
on the PUF so that the total time
\begin{equation}
\Delta t_t = \Delta t_r \times N
\end{equation}
to read them all out is 
much longer than the time $\Delta$t$_a$ during which an attacker 
possesses the PUF.
$\Delta t_r$ is the read-out time for one C -- S pair.
The maximal fraction of pairs the attacker can read-out is
then $\Delta$t$_a$/$\Delta t_t$ = L$_{bf}$. L$_{bf}$ is the security
level against mathematical-cloning brute force attacks.
\\
Pappu's optical
PUF\cite{pappu}, the arbiter PUF\cite{gassend04} and 
nearly all other current PUFs are MRT-PUFs\footnote{The only exception
are ``PUFs'' with only one challenge which we propose
to call only ``POKs'' in the future, see section \ref{pok}.}.
These constructions are valid PUFs according to
our definition because their values of PF$_2$ changes with the challenge.
\\
However, many of the existing
PUFs are insecure according to our definition 
because R\"uhrmair et al.\cite{ruehr} succeeded to employ
machine-learning methods
that allow to infer PF$_2$(PF$_1$())
from a small fraction of all C -- R
for which only short $\Delta$t$_a$ is necessary\cite{ruehr}. 
Because all C -- S pairs can be thus predicted, the security level against
machine-learning attacks L$_{ml}$ = 1 which is ``not negligible''
in general, i.e. the PUF must be considered mathematically 
clonable according
to PUF-security definition 2.
\\
The exact form of PF() depends on the detailed architecture of
the MRT PUF. In general
MRT PUFs can be hardened against mathematical
cloning if their PF$_2$(PF$_1$) fulfills the following demands:
\\
{\bf Security requirements for the MRT-PUF}
\begin{itemize}
\item
{\it N must satisfy: $N \geq L^{-1} (\Delta t_a/\Delta t_r $)}
\item
{\it Suppose PF$_2$(PF$_1$(C$_n$)) = S$_{n}$ with n = 1...N where
both C$_n$ and S$_{n}$ contain $\ell$ bits. Then
the combined information content (entropy) I of all C$_n$ and
S$_{n}$ must satisfy: I $\geq$ 2 N $\ell$}
\item
{\it The set of challenges to be used in operation must not
be contained in any form in the PUF.}
\item
{\it The lengths of the challenge $\ell$
and response $\ell_S$ must both fulfill:
$\ell$,$\ell_S$ $\geq$ log$_2$(N).}
\end{itemize}
The first condition expresses that 
to prevent brute
force mathematical-cloning attacks
the number of stored C -- R pairs N must be extremely large 
if L = 10$^{-15}$, (see
section \ref{securepuf} on the choice of L). With representative
values of $\Delta t_a$ = 1 day and $\Delta t_r$ = 1 second the required N
would be on the order of 10$^{20}$ which is exponentially larger
than e.g. storable in common data storage devices of much larger
size than a typical PUF. This is the sense in which a secure
MRT-PUF requires the storage of an ``exponentially large'' secret.
The second condition expresses that
in order to reliably ward
successful machine-learning attacks
PF$_2$ must be just an ordered list of C -- S
pairs with random values 
that cannot be represented in any more compact
form. The third requirement prevents an attack in which
only the set of challenges to be used in the field operation of a PUF 
(which is much smaller than $\mathfrak{M}$ in secure MRT PUFs) are extracted in an attack.
The  fourth constraint is necessary to avoid a decrease in the
the effective L.

\subsection{``Erasure Upon Readout'' PUFs -- Quantum PUFs}
\label{quant}
Consider a PUF with only a single C -- S pair
foreseen by the system architecture. Because
there is at least one other non-foreseen C,
there are then at least two possible C.
A novel PUF security mechanism requires the following:
\\
{\bf Security requirements for ``Erasure Upon Readout'' (EUR) PUF}
\begin{itemize}
\item
{\it The correct S is returned if the challenge C is correct
(i.e. the one foreseen by the PUF's architecture)
and S is erased and returns a random value if it is not.}  
\item
{\it The length of the challenge $\ell$ and response $\ell_S$ must both
fulfill $\ell$,$\ell_S$ $\geq$ log$_2$(1/L).}
\item
{\it The set of challenges to be used in operation must not
be contained in any form in the PUF.}
\end{itemize}
EUR PUFs can fulfill the PUF-definition 1
if they are non-constant PFs that
are deterministic for the foreseen set of
challenges.
For EUR PUFs - completely opposite to the MRT case (see
section \ref{mrt}) - the total number of challenges ``N'' can remain as small
as 2 but still be secure because by way of the second and third security requirement
the probability to guess the correct challenge is only L and challenging
with the wrong challenge will erase S by the first requirement.
N can be chosen to as many different challenges as are actually
needed during the practical deployment of the PUFs.
\\
The only concrete ``Erasure Upon Readout'' architecture proposed up to now,
is Wiesner's ``quantum money'' and ``quantum unforgeable subway token''\cite{wiesner,bennett}
that can be described as an electronic hardware device running a challenge - response protocol 
(such a kind of ``money'' or ``token'' has to be) 
and that fulfill our definition 1 of a PUF.
In such a ``quantum-PUF'' 
the secret information 
consists of $\ell$ quantum-mechanical two-state systems
(``qubits'') that are prepared
either in one of the two quantum mechanical so called
``Fock'' states $|0\rangle$
or $|1\rangle$ (base $\#$ 0)
or in either one of the states $1 \over \sqrt{2}$ ($|0\rangle$ + $|1\rangle$)
or $1 \over \sqrt{2}$ ($|0\rangle$ - $|1\rangle$) (base $\#$ 1).
$|0\rangle$ and $1 \over \sqrt{2}$ ($|0\rangle$ + $|1\rangle$) encode a ``0'' secret bit and 
$|1\rangle$ and  $1 \over \sqrt{2}$ ($|0\rangle$ - $|1\rangle$) encode a
``1'' secret bit. The
challenge bits indicate the correct chosen measurement bases.
The raw secret S$_r$ is encoded with the choice of the state within a
chosen basis according to the rule stated above.
\\
In order to decode or copy S$_r$, it is necessary to know
in which of the two bases $\#$ 0 or $\#$ 1 the $\ell$ qubits for one
challenge were
prepared. If a qubit is read out in a wrong base, 
the laws of quantum mechanics determine that the read-out
result is a perfect random number and additional read out
attempts will again yield this random number, rather than the
original, correct number. The physical function PF of
the quantum-PUF is thus given as:
\\
{\bf Quantum-EUR PF$_1$()}:
\begin{itemize}
\item
First read-out:
\\
PF$_1$(correct base bit) = correct bit of S$_r$
\\
PF$_1$(incorrect base bit) = random bit.
\item
Any further read-out in the same base:
\\
PF$_1$() = same bit as in first read-out
\end{itemize}
Evidently in the first read-out PF$_1$ is not constant
and deterministic for the foreseen C i.e. a quantum-PUF fulfills
definition 1.
Reading out a C -- S pair more than once is possible, but
after the first read-out, the information is no longer secure
because the qubits are no longer in a 
quantum-mechanical superposition of states.
\\
In the most simple case without any read-out errors or inefficiencies
(so that no further processing is done PF$_2$(S$_r$)=
S$_r$)
and implementation mistakes (an assumption that will be
difficult to fulfill \cite{scarani})
the only potentially successful attack is to guess the challenge.
On average, for half of the bits the guess will be correct
and the correct corresponding bits of S$_r$ will be output.
For the other half the probability to get the correct
output bit is 1/2. The total probability to get a correct
output bit of S$_r$ is therefore 0.75 
and L$_a$ = $({3 \over 4})^{-\ell}$, which is the absolute
(i.e. not only mathematical-cloning brute force) security level
of a quantum-PUF against this attack.
E.g. with a secret S$_r$ consisting of 128 qubits,
L $<$ 10$^{-15}$ thus fulfilling the criterion for 
a secure PUF with Borel's estimate for an upper
bound on L (see section \ref{securepuf}).
Wiesner's quantum money, interpreted as a ``quantum PUF'', thus proves that an absolutely
unclonable PUF
is not in contradiction to the laws of physics.
\\
The use of quantum-PUFs for authentication is beyond the reach of
current technology because
qubits  
are unavoidably read out 
on very short timescales (presently qubits cannot be isolated for 
longer than milliseconds\cite{qubit}) by interactions with their environment. 
As explained above, quantum-PUFs are no longer
secure after read-out.
Quantum cryptography\cite{scarani} can be described as sending a quantum-PUF
in the form of a chain of photons in order to distribute its secret S 
for use as cryptographic key.
In the laboratory such a ``light-field'' PUF remains in the
initially prepared coherent state for
no longer than about a millisecond.

\section{Discussion}
\label{discu}
The protection of secrets in hardware devices that need
to access these secrets in their normal operation 
- a necessary condition for any authentication procedure -
cannot be implemented with methods of mathematical cryptography
alone. Some physical protection mechanism is needed.
The conventional tamper resistance mechanisms (employed
in CUFs see section \ref{cuf})
rely on protecting the memory
with physical barriers.  
CUFs withstand known, vigorous
direct attacks typically for not longer than a few months\cite{tarnovsky}.
We showed that PUFs are a {\it qualitatively} novel alternative.
The secret is protected by the absence of information 
from the device of where 
of where the challenge is stored. In CUFs and POKs this
information must exist on the device because otherwise
the response cannot be evaluated,
even if it is
protected by direct, physical barriers. 
Thereby PUFs protect the secret by a novel genuine primitive
of physical cryptography. 
The possibility of realizing PUFs 
based on the principles of quantum mechanics demonstrates that
in principle the laws of physics allow to construct absolutely secure PUFs.
This situation motivates
more security-related physics research on 
unclonable quantum-PUF and MRT-PUF, to invent
entirely new PUF construction principles.
The real PUF promise are
PUFs that withstand any known, practical attack, period,
i.e. provide a level of authenticity protection similar to the one provided
by mathematical cryptography for confidentiality.
\\
In the future
PUFs will probably authenticate hardware
devices. If Alice knows the C -- S$_r$ pairs of a PUF
she gave to Bob (e.g. from the
designer of the PUF) she can publicly broadcast a challenge
and be sure that the correct response S can
only be created on Bob's original PUF.
Therefore effectively PUFs
allow the remote distribution of authenticated secret entropy (the
S for Bob) 
via sending the challenges (the C chosen and sent by Alice) over standard channels.
These entropy could ``update'' the secrets in conventional
unclonable functions. In this way existing architectures based on
CUFs could be augmented by PUFs without the need for 
a completely new PUF security architecture.
\\
{\bf Acknowledgements.} We thank R. Breithaupt, U. Gebhardt, M. Ullmann, C. Wieschebrink
and anonymous referees at the TrustED 2011
and PILATES 2012 workshops for
helpful discussion and criticism on earlier versions of this manuscript.

\end{document}